\documentstyle[11pt,psfig]{article}    
\newcommand{\be}{\begin{equation}}
\newcommand{\ee}{\end{equation}}
\newcommand{\bea}{\begin{eqnarray}}
\newcommand{\eea}{\end{eqnarray}}
\begin{document}

\setlength{\baselineskip}{22pt}   

\begin{center}
{\LARGE \bf Excitation of Color Degrees of Freedom of Nuclear Matter and 
            {\boldmath $J/\psi$} Suppression}\\[5mm]
{J\"org H\"ufner$^{a,b}$, Boris Z. Kopeliovich$^{b,c}$ and 
 Alberto Polleri$^a$}\\[4mm]
{\small{\it $^a$ Institut f\"ur Theoretische Physik der Universit\"at, 
            Philosophenweg 19, D-69120 Heidelberg, Germany.} \\[1mm]
       {\it $^b$ Max Planck Institut f\"ur Kernphysik, Postfach 103980,
            D-69029 Heidelberg, Germany.} \\[1mm]
       {\it $^c$ Joint Institute for Nuclear Research, Dubna, 
            141980 Moscow Region, Russia.} \\[5mm]}
\end{center}

\begin{abstract}
\setlength{\baselineskip}{22pt}   
In high energy nuclear collisions, the conventional Glauber model is commonly used to
evaluate the contribution to $J/\psi$ suppression originating from the inelastic 
interaction with colorless bound nucleons. This requires an effective value for the
$J/\psi$-nucleon absorption cross section which is larger than theoretically
expected. On the other hand, multiple nucleon-nucleon collisions mediated by color
exchange interactions, excite their color degrees of freedom. We investigate the 
importance of this effect and find that these excited states provide a larger cross 
section for $J/\psi$ absorption. We conclude that the related corrections are 
important to explain the effective value extrapolated from experiment.
\end{abstract}

\vspace{0.2in}

\setlength{\baselineskip}{22pt}   


\section{Introduction}

The origin of the anomalous behavior of the $J/\psi$ production
cross section, as measured in Pb+Pb collisions at the CERN-SPS is 
still debated and several competing interpretations have so far been 
proposed. For a review on the subject see  \cite{GH,V}. 

In the early stage of a nucleus-nucleus ($AB$) collision, the production mechanism
of a charmonium is controlled by four length scales. They are the 
coherence length $l_c \leq 2\,E_{J/\psi}/m^2_{J/\psi}$, the formation length $l_f 
\simeq 2\,E_{J/\psi}/(m^2_{\psi'}-m^2_{J/\psi})$, the mean interparticle distance 
$\lambda \simeq 2$ fm in a nucleus and the mean nuclear size $R_{A,B} \simeq 5-6$ fm 
for large nuclei. Their
interplay is crucial for the correct theoretical formulation of the problem and
understanding the nature of early stage suppression. In the kinematic regime of the
NA38/50 experiment at the CERN-SPS, which detects charmonia with $p_{lab} = 50$ GeV,
one has $l_c \leq 2$ fm and $l_f \simeq 5$ fm. Therefore, since $l_c$ is shorter than 
the mean nucleon separation, one can assume that the colorless $c \overline c$ wave 
packet is produced in 
elementary nucleon-nucleon ($NN$) interactions. On the other hand, the formation
length is of the order of the nuclear size and the charmonium takes a finite path
in the nucleus before being fully formed. The effect of these length scales on 
hadronic processes in a nuclear environment has been extensively studied 
\cite{KZ91,HKN96,HK96,HKZ} and experimentally observed in electroproduction of $\rho$
mesons off nuclei \cite{A99}. 
Nevertheless, $J/\psi$ suppression is mostly treated 
within the Glauber model, with a constant effective cross section for 
absorption on bound, non-interacting nucleons. Only recently \cite{HK96,HeHK}, the 
effect caused by the formation length was taken into account in the present context.
Notice that while 
studying the SPS data it is justified to neglect the contribution of the coherence 
length, for RHIC data it will not be so, being $l_f \gg l_c \gg R_{A,B}$, and a new 
regime will open up. For the present discussion on SPS data, the relevant scales for
$J/\psi$ absorption are $l_f$ and $R_{A,B}$, governing the suppression at the early
stage.
         
After the early stage, governed by $NN$ collisions, an excited medium with high 
energy density is left behind. It 
is not yet clear how to describe it, but several points of view have been put 
forward. It is possible that the properties of the excited medium are similar to
those of a plasma of de-confined quarks and gluons. This was indeed the original 
motivation for making $J/\psi$ suppression so interesting \cite{MS86}.
It has been argued that such a plasma is very opaque to charmonium, therefore 
preventing its final observation as a bound state \cite{BO96,KLNS97}. 
In the last stage of the reaction, part of the large number of produced hadrons,
the so called co-movers, is possibly also responsible for suppression 
\cite{GV97,AC98,ACF99}. 

The present work intends to study the early stage nuclear absorption of $J/\psi$.
It is common practice to approach the problem assigning a constant value to the
$J/\psi$-nucleon absorption cross section. This quantity is then considered as 
an effective parameter, fitted in order to reproduce the production data 
in proton-nucleus (pA) collisions. Our goal is to improve the understanding of the 
physical phenomena behind early stage absorption, neglecting for the time being
other possible effects and therefore not necessarily being able to describe of the
full set of available data.

Following a known interpretation of the dynamics of $NN$ collisions, we ascribe the 
main contribution to the inelastic cross section to color exchange processes. This 
assumption is well supported by high-energy hadron collision data, showing a plateau
in the rapidity distribution of produced particles and indicating that the real part
of the scattering 
amplitude is small with respect to the imaginary one. Via color exchange,
the 3-quark systems constituting the colliding nucleons become
color octet states. It was already mentioned in \cite{HK98} that each 
colliding system is colored, although the possible effects of this were neglected 
for simplicity. Color exchange may also be accompanied by gluon radiation. While
the contribution of radiated gluons to $J/\psi$ absorption was treated 
in \cite{HK98,HuHK00}, here we focus on the first part. Having realized that
color-singlet nucleons are found in color octet states after an elementary 
collision, one can expect that repeated scatterings lead to multiple color 
exchanges. One
therefore has the possibility to excite the color degrees of freedom of the
3-quark system also to the decuplet state. Higher color representations can be 
obtained when radiated gluons are also present, but here we restrict our study on
the color dynamics of the constituent quarks of a nucleon.

In the next section we calculate the cross sections for the interaction of 
charmonium with a colored nucleon. We find an increase of $35\,\%$ when the 
3-quark system is in octet state, with respect to the
case when it is a color singlet, while with the decuplet the increase is of 
$70\,\%$. We analyze and discuss in detail $pA$ collisions in the third section. 
Finally, we consider the $AB$ case: the effect of multiple scattering is 
studied and then applied to calculate integrated production cross sections. We find
an increased absorption, although the experimental suppression in the Pb+Pb cross
section cannot be explained by the present 
treatment. The concluding section critically examines the results and gives an 
outlook on possible further improvements. Throughout the paper we indicate with 
$\Psi$ the effective charmonium state, composed by a mixture of 
$J/\psi$, $\psi'$ and $\chi$.

\section{Interaction of Charmonium with Colored Nucleons}

The main feature we want to stress is the 
property of the color exchange interaction for the high energy collision 
amplitude. Repeated collisions allow a nucleon to be in any
color SU(3) representation ${\bf 3} \otimes {\bf 3} \otimes {\bf 3} = 
{\bf 1} \oplus {\bf 8}_{A} \oplus {\bf 8}_{S} \oplus {\bf 10}$. To calculate
the value of the inelastic cross section for $\Psi$ on a nucleon
which has undergone multiple scattering, we use the two-gluon exchange model of 
Low and Nussinov \cite{LN75,GS77}. Although oversimplified, the model
incorporates important properties of high-energy hadronic cross sections such as 
(approximate) energy independence and scaling of the cross sections according to
the mean square radius of hadrons.
One can write at once the absorption cross section, making use of the 
optical theorem, relating it to the forward elastic amplitude. This is convenient,
since the latter amplitude, in the two gluon exchange approximation, is purely 
imaginary. One obtains
\be
\sigma^{abs}_{\Psi N} = \int d^2\vec{r} \ \,
\mbox{{\Large $\langle$} $\!\!\Psi,N_n\!\!$ {\Large $|$}}
\, \mbox{\Large $[$}\, \sum_{i,j} V(\vec{b} + \vec{x}_i - \vec{y}_j) 
\, \mbox{\Large $]$}^2 
\mbox{{\Large $|$} $\!\!\Psi,N_n\!\!$ {\Large $\rangle$}}.
\label{crossfull}
\ee
\begin{figure}[t]
\centerline{\psfig{figure=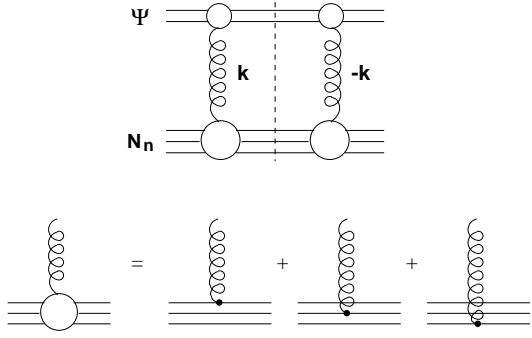,width=7cm,angle=-90}} 
\protect\caption{Top: Diagram illustrating the meson-nucleon elastic amplitude. The
dashed line indicates the unitarity cut. Bottom: Vertex $\Gamma$ between a gluon 
and the three-quark system of a nucleon.}
\label{diagrams}
\end{figure}
It is a diagonal matrix element between a product state of charmonium $\Psi$ and 
a nucleon $N_n$ in its color representation $n$. The 
variable $\vec{b}$ is the impact parameter between the scattering hadrons.
The transverse coordinates $\vec{y}_{1,2,3}$ refer the quarks in the nucleon,
while $\vec{x}_{1,2}$ are those of $\Psi$.
In the square brackets one has the transverse part of the one-gluon 
exchange potential. We define its Fourier transform as
\be
V(\vec{z}) = \int \! \frac{d^2\vec{k}}{(2\pi)^2} 
\ \mbox{\large$\frac{1}{2}$} g_s \lambda^a_{(i)}
\ \frac{\exp(\,i\, \vec{k}\,\vec{z}\,)}
{k^2 + \mu^2}\ \mbox{\large$\frac{1}{2}$} g_s \lambda^a_{(j)},
\label{gluepot}
\ee
where $g_s$ is the quark-gluon QCD coupling and $\mu$ is an effective gluon mass,
introduced to incorporate confinement.
The index in parenthesis labeling the Gell-Mann matrices indicates to which 
quark the vertex is attached. Light-cone wave functions are used for $\Psi$ and
$N_n$ in the expectation value (\ref{crossfull}). For the nucleon, we assume
that the transverse positions of the quarks in the nucleon are frozen during the 
early stage of the collision, and therefore
use the same spatial distribution both for singlet and colored states.

\begin{figure}[t]
\centerline{\psfig{figure=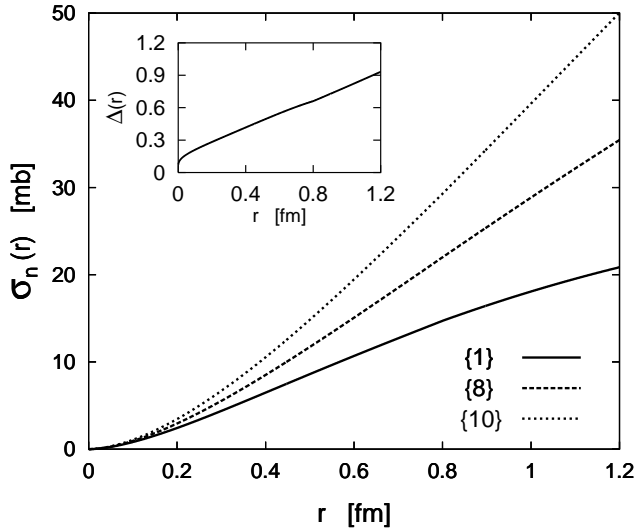,width=8.5cm,angle=-90}} 
\protect\caption{Dipole cross section for different color states as a function
of the quark separation $r$. Box: The quantity $\Delta(r)$ which governs the 
dependence on color.}
\label{FIGDIP}
\end{figure}

Inserting in eq.$\,$(\ref{crossfull}) the expression for the gluon potential
given in eq.$\,$(\ref{gluepot}) and performing the impact parameter integration,
one arrives at the relation
\bea
\sigma^{abs}_{\Psi N_n} & = & \int d^2\vec{k} \ \frac{1}{(k^2 + \mu^2)^2}
\ \mbox{{\large $\langle$} $\!\!\Psi\!\!$ {\large $|$}}
\, 1 - \exp(\,i\, \vec{k}\,\vec{x}_{12}\,)
\, \mbox{{\large $|$} $\!\!\Psi\!\!$ {\large $\rangle$}} 
\nonumber \\
& & \ \ \ \ \ \ \ \ \ \ \ \times
\, \mbox{{\large $\langle$} $\!\!N_n\!\!$ {\large $|$}}
\,\Gamma^a(k) \,\Gamma^{\dagger\,a}(k)
\,\mbox{{\large $|$} $\!\!N_n\!\!$ {\large $\rangle$}} \,,
\label{colorcross}
\eea
illustrated in the top part of FIG.~\ref{diagrams}.
The first factor in the integrand is the squared gluon propagator,
the second contains the phase shift due to the momentum transfer $\vec k$ and is 
responsible, by the way, for color transparency,
while the last factor is the interesting part of the present calculation. It is 
the product of two vertices
between three quarks and a gluon, averaged with the wave function of the 
colored nucleon. The vertex, illustrated in the bottom part of 
FIG.~\ref{diagrams}, is
\bea 
\Gamma^a(k) & = & \mbox{\large $\frac{16}{3}$} \, \alpha_s \, \left[
\, (\lambda^a_{(1)})^i_{i'}\,\delta^j_{j'}\,\delta^k_{k'}\,
\exp(\,i\, \vec{k}\,\vec{y}_1\,) \right. \nonumber \\
& & \ \ \ \ \ \ \ \, + \ \delta^i_{i'}\,(\lambda^a_{(2)})^j_{j'}\,\delta^k_{k'}\,
\exp(\,i\, \vec{k}\,\vec{y}_2\,) \nonumber \\
& & \ \ \ \ \ \ \ \, + \left .\delta^i_{i'}\,\delta^j_{j'}
\,(\lambda^a_{(3)})^k_{k'}\, \exp(\,i\, \vec{k}\,\vec{y}_3\,)\, \right] .
\eea
In order to average the product of the two vertices, one needs to specify the
structure of the nucleon wave function. The color part of the wave 
function is the relevant one, while the spatial part is chosen
to be equal for all color multiplets. We (anti)symmetrize quarks 1 and 2 in order
to construct the mixed-symmetry octets, therefore use
\be
\phi^n_{ijk} = 
\left\{
\begin{tabular}{cc} 
$\mbox{\large $\frac{1}{\sqrt{6}}$}\,\epsilon_{ijk}$ & $\ \ \{\mbox{\bf 1}\}$\,, \\
$\mbox{\large $\frac{1}{\sqrt{2}}$}\,\left[\delta^r_k\,\epsilon_{ijs}
- \mbox{\large $\frac{1}{3}$}\,\delta^r_s\,\epsilon_{ijk}\right]$ & 
$\ \ \{\mbox{\bf 8}_A\}$\,, \\
$\mbox{\large $\frac{1}{\sqrt{6}}$}\,\left[\delta^r_i\,\epsilon_{jks}
+ \delta^r_j\,\epsilon_{iks}\right]$ & $\ \ \{\mbox{\bf 8}_S\}$\,, \\
$\mbox{\large $\frac{1}{6}$}\,\left[\delta^r_i\,\delta^s_j\,\delta^t_k
\ + \ \mbox{perm. of $r,s,t$}\,\right]$ & $\ \ \{\mbox{\bf 10}\}$\,. \\
\end{tabular}
\right.
\ee
Because of the $SU(3)$ structure, the different color multiplets have different
number of indices. For simplicity, we label with $n$ the different multiplets.
The factor given by the angular brackets in eq.$\,$(\ref{colorcross}) can now be
evaluated. It involves the contraction of all the indices, leading to
\be
\mbox{{\large $\langle$} $\!\!N_n\!\!$ {\large $|$}}
\Gamma^a\,\Gamma^{\dagger\,a} 
\mbox{{\large $|$} $\!\!N_n\!\!$ {\large $\rangle$}}
= \mbox{{\large $\langle$} $\!\!N\!\!$ {\large $|$}}
\sum_{i < j}\, 1 - c_n \exp(\,i\, \vec{k}\,\vec{y}_{\,ij}\,)
\mbox{{\large $|$} $\!\!N\!\!$ {\large $\rangle$}}.
\label{colourcrossfin}
\ee
The effect of having different color states is all contained in the
coefficients $c_n$, listed in the second column of TABLE~\ref{psinuc}. It turns
out that matrix elements with $n = {\bf 8}_{A},\,{\bf 8}_{S}$ are identical.
The values of $c_n$ found, for $n = {\bf 8},\,{\bf 10}$, lead to larger absorption
cross sections as compared to the singlet case.
To see this it is useful to define the two-quark 
form factor $F^{2q}_N(k^2) = \mbox{{\large $\langle$} $\!\!N\!\!$ {\large $|$}}
\exp(\,i\,\vec{k}\,\vec{x}_{\,ij}\,)
\mbox{{\large $|$} $\!\!N\!\!$ {\large $\rangle$}}$
for the nucleon. Since, within our approximation, the latter
does not depend on the colour multiplet, we drop the label $n$. Then we can
rewrite eq.$\,$(\ref{colorcross}) as 
\be
\sigma^{abs}_{\Psi N_n} = 
\mbox{{\large $\langle$} $\!\!\Psi\!\!$ {\large $|$}}
\,\sigma_n(r)\,
\mbox{{\large $|$} $\!\!\Psi\!\!$ {\large $\rangle$}}\,,
\label{abscross}
\ee
where the function in the brackets is
\bea
\sigma_n(r) & = & \mbox{\large $\frac{16}{3}$} \ \alpha_s^2 
\int d^2\vec{k} \ \frac{1}{(k^2 + \mu^2)^2}
\  [ 1 - \exp(\,i\, \vec{k}\,\vec{r}\,)] \nonumber \\
& & \ \ \ \ \ \ \ \ \ \ \ \ \ \ \ \ \times\, [1 - c_n F^{2q}_N(k^2)]\,,
\label{dipole}
\eea
exhibiting a similar structure to that of the usual dipole-nucleon cross section, 
but with the important difference carried by the coefficients $c_n$.
\begin{table}[b]
\caption{Results of the calculation of the meson-colored nucleon cross sections.}
\begin{center}
\begin{tabular}{|c||c|c|c|} \hline
Multiplet $\ $ \{n\} & $\ \ c_n\ \ $ & $\ \sigma_{\psi N_n}$ [mb] $\ $ 
	& $\ \ \delta_n \ \ $\\ \hline\hline
Singlet  $\ \ \ \;$ \{{\bf 1}\}   &   $\ 1$    & 5.8 &  1.0   \\ \hline
Octet  $\ \ \ \ \ $ \{{\bf 8}\}   & $\, -1/4\ $ & 7.8 &  1.35   \\ \hline
$\ $ Decuplet $\ $\{{\bf 10}\} $\ $ &  $\ \,1/2$   & 9.9 &  1.7   \\ \hline
\end{tabular}
\end{center}
\label{psinuc}
\end{table}
Notice that with $c_n \neq 1$ and $\mu = 0$ the integral in eq.$\,$(\ref{dipole})
is infrared divergent. Therefore, the cut-off $\mu$, of the order of the inverse
confinement radius, is important. With $\mu \neq 0$ it holds that
 $\sigma_n(r) \rightarrow 0$ for $r \rightarrow 0$, so 
that one still preserves color transparency, even with a colored nucleon. This is
indeed a property of the singlet meson. In our work we are not primarily interested
in the absolute values of $\sigma^{abs}_{\Psi N_n}$ but rather on its dependence
on the color state $n$ with respect to the singlet value. Yet we estimated the 
absolute values choosing $\alpha_s = 0.6$ and $\mu = 140$ MeV. Within a harmonic
oscillator model of the nucleon, one finds that $F^{2q}_N(k^2) = F^{em}_N(3k^2)$,
the electromagnetic form factor. We assume that this property is general and
take the dipole form
\be
F^{em}_N(Q^2) = \left(\frac{{\tilde \lambda}^2}{{\tilde \lambda}^2 + Q^2}\right)^2
\mathop\simeq_{Q^2 \rightarrow 0} 1 - \frac{1}{6}\, Q^2\,R_p^2\,,
\label{PROTFF}
\ee
where the last approximate equality allows to fix the parameter ${\tilde \lambda}^2
= 12 / R^2_p$, with $R_p = 0.8$ fm. For convenience we rescale ${\tilde \lambda}^2 = 
3 \lambda^2$, so that $\lambda = 2 / R_p$.
With the above definitions, the generalized dipole-nucleon cross section can be 
computed analytically in terms of modified Bessel functions $K_0$ and $K_1$,
with the result
\bea
\sigma_n(r) & = & \frac{16 \pi \alpha_s^2}{3 \lambda^2} 
\left\{ 
\frac{1 - D_2^n(\xi)}{\xi^2}\,H(\mu r) - D_2^n(\xi)\,H(\lambda r) 
\right.
\nonumber \\
& & 
- \, D_3^n(\xi)\,\left[ \log(\xi) + K_0(\mu r) - K_0(\lambda r) 
\right] 
\mbox{\large $\}$}
\,,
\eea
being $D_2^n(\xi) = c_n / (1 - \xi^2)^2$, $D_3^n(\xi) = 4 c_n / (1 - \xi^2)^3$,
$\xi = \mu / \lambda$ and $H(r) = 1 - z K_1(r)$. 
The results of the calculation, illustrated in FIG.~\ref{FIGDIP}, 
shows the modified dipole-nucleon cross section for the three color
states of the nucleon. It largely increases when the dipole scatters off a
colored object. One can get more insight in the effect of the different color
states rewriting eq.$\,$(\ref{dipole}) in the form
\be
\sigma_n(r) = \sigma(r) \left [\, 1 + (1 - c_n)\,\Delta(r) \,\right]\,,
\label{deltadip}
\ee
where $\sigma(r)$ is the usual dipole cross section on a color singlet nucleon
and the function 
$\Delta(r)$ is found to vary appreciably with $r$. Its functional dependence is
given in the small box in Fig. \ref{FIGDIP}. To calculate the inelastic cross 
section according to eq.$\,$(\ref{abscross}),
we must average $\sigma_n$ with the meson wave function. Therefore, the dependence
of $\Delta$ on $r$ shows that the effect of having colored nucleons is stronger for
larger size mesons.  Recent calculations \cite{HIKT}, based on realistic phenomenology
for the dipole cross section and charmonium wave function, gave values for
cross sections of various charmonia on a nucleon. At center of mass energy $\sqrt{s}
= 10$ GeV, it was found that $\sigma_{J/\psi N} = 3.6$ mb, $\sigma_{\psi' N} = 
12.2$ mb and
$\sigma_{\chi N} = 9.1$ mb. With a composition of $\Psi$ of 52-60$\%$ due to 
$J/\psi$, 8-10$\%$ due to $\psi'$ and 32-40$\%$ due to $\chi$, a weighted value 
of 5.8 mb was obtained. 
In the present case we evaluate $\sigma^{abs}_{\Psi N_n}$ with
a Gaussian wave function having root mean square transverse separation 
$\langle r^2 \rangle^{1/2} = 
0.25$ fm, in order to account for the different states ($J/\psi$, $\psi'$, $\chi$) 
implicitly included in $\Psi$ and therefore adjust the singlet cross section to the 
quoted 5.8 mb.
The values found are given in the third column of TABLE~\ref{psinuc}. In order to
minimize the model dependence, we factored out from the colored cross sections the
amount corresponding to the singlet case, therefore writing
$\sigma^{abs}_{\psi N_n} = \sigma^{abs}_{\psi N} \ \delta_n$.
The values for $\delta_n$ are given in the last column of TABLE~\ref{psinuc}.
They constitute the important result of this calculation ,
showing that $\Psi$ scatters with an color octet nucleon with a $35\,\%$ larger
cross section with respect to a singlet nucleon, while with the decuplet the
increase is even of $70\,\%$. The large values found now require a careful analysis
to establish the validity of this newly proposed absorption mechanism.

\section{Analysis of proton-nucleus data: is the color mechanism already at work ?}

Since the absorption pattern observed in $pA$ collisions is usually 
considered to be the baseline for the $AB$ case, it is necessary to 
examine it in some detail in the light of the possibility of a color 
mechanism for charmonium absorption.

\begin{figure}[t]
\centerline{\psfig{figure=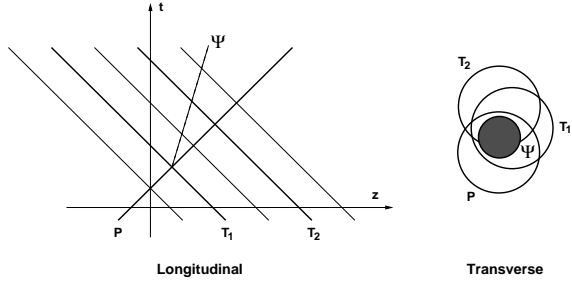,width=7.5cm,angle=-90}} 
\protect\caption{Left: Longitudinal space-time representation of charmonium
production in a $pA$ collision, in the c.m. frame of a $NN$ collision. Right: 
Illustration
of the overlap between projectile (P), parent target (T$_1$) and nucleon with 
which $\Psi$ scatters inelastically (T$_2$).}
\label{protnuc}
\end{figure}

Consider the illustration given in FIG~\ref{protnuc}. There one can see
the longitudinal and transverse geometry of the collision between a proton and 
the row of target nucleons with which it collides. On the left hand side one sees
the projectile P which collides 
with the target nucleon T$_1$, producing a would be $\Psi$ meson. Successively
the projectile interacts with  the target nucleon T$_2$ which, then, collides
with the $\Psi$ meson. According to our initial argument, each nucleon-nucleon 
collision leaves the nucleons in a color octet state. As illustrated
on the right hand picture, since $\Psi$ is produced by P, it is overlapping with
it. If then the charmonium is struck by T$_2$, P and T$_2$ must
also overlap, indicating that they have previously interacted. Therefore, in a
$pA$ collision all the nucleons with which $\Psi$ has an interaction
are color octets. This means that the effective absorption cross 
section normally used in Glauber model calculations in order to reproduce $pA$ data,
should be the cross section with a color octet nucleon. On the other hand, another
simple geometrical argument shows that this might not be the full truth. 
In fact, in order to claim that $\Psi$ interacts with a colored
nucleon, the charmonium state itself is required to be well separated from the parent
projectile 3-quark system, preceeding $\Psi$ in propagation through the nucleus.
This takes a finite time, since the charmonium is produced
inside the projectile and has a smaller velocity. In the rest frame of the 
nucleus, the difference in velocities between $\Psi$ and the parent 3-quark system is
$\Delta v \simeq m_\Psi^2 / 2 E_\Psi^2 - m_p^2 / 2 E_p^2$, being $v = \sqrt{1 -
m^2/E^2} \simeq 1 - m^2 / 2 E^2$. Therefore, one requires a finite length
$L = R_p / \Delta v$, with $R_p = R_p^{(0)} / \gamma_p$, in order to have the 
charmonium separated from the projectile. In the following we focus on the kinematics
of the NA38/50 experiments at the CERN-SPS, for
charmonia with laboratory momentum $p_\Psi = 50$ GeV and consider projectile 
protons with at least $E_p = 200$ GeV. Then, since $\gamma_p = E_p/m_p$,
taking $R_p^{(0)} = 0.8$ fm we obtain $L = 2$ fm. The physical meaning of $L$ is 
the following: After its creation, the charmonium has to travel a length $L$ (in 
the target rest frame) before one can safely assume that the next interaction 
occurs with a colored nucleon. Since the mean length which $\Psi$ travels inside 
a nucleus is $2/3\,R_A$, the retardation length $L$ reduces the effect of colored 
nucleons considerably. It destroys it in light nuclei with $A \leq 50$, and
it reduces it to about 50$\%$ in heavy nuclei.

Now, before we analyze the data for charmonium production in $pA$ collisions, we
discuss the importance of charmonium formation time, introduced in a  
quantum-mechanical way in \cite{KZ91,HK96}. Within a simplified version of a 
two-channel model \cite{HeHK},
the $\Psi N$ absorption cross section evolves in the eigentime $\tau$ of $\Psi$ as
\be
\sigma^{abs}_{\Psi N}(\tau) = \sigma^{in}_{\Psi N} + (\sigma^{(0)}
- \sigma^{in}_{\Psi N})\,\cos(\tau/\tau_f)\,,
\label{crossft}
\ee
where $\tau_f = (m_{\psi'} - m_{J\psi})^{-1} = 0.3$ fm is the charmonium formation
time, $\sigma^{in}_{\Psi N}$ is the inelastic $\Psi N$ cross section and
$\sigma^{(0)} = \sigma^{abs}_{\Psi N}(\tau = 0)$ is the cross section 
of the so called ``pre-meson'' on a 
nucleon. As mentioned in the last section, calculations give 
$\sigma^{in}_{\Psi N} = 5.8$ mb \cite{HIKT}. Unfortunately $\sigma^{(0)}$ has
not yet been calculated. We therefore use the value 2.7 mb from \cite{HeHK}. 

Making use of eq.$\,$(\ref{crossft}), one can then write the 
expression for $\Psi$ production in $pA$ collisions, normalized to one nucleon, as
\bea
\sigma_\Psi^{pA} & \!\!=\!\! & \,\sigma_\Psi^{NN} \frac{1}{A} 
\int d^2\vec{b}\,dz \label{glaufirst} \\
 & & \times\,\left[ 1 - \int_z^{+\infty}\!\!\!dz'
\ \sigma^{abs}_{\Psi N}\mbox{\large$($}\mbox{\large $\frac{z'-z}{\gamma_\Psi}$}
\mbox{\large$)$}
\ \rho_A(z,\vec{b})/A
\right]^{A-1}\!\!\!. \nonumber
\eea
We now want to compare this expression with the conventional one obtained in the
Glauber model, containing
only an effective absorption cross section, therefore making the replacement
$\sigma^{abs}_{\Psi N}(\tau) \rightarrow \sigma^{eff}_{\Psi N}$. In this way one
can perform the $z$-integration and obtain the well known formula
\be
\sigma_\Psi^{pA} = \,\sigma_\Psi^{NN} \int d^2\vec{b}
\ \,\frac{1 - \left[1 - \sigma^{eff}_{\Psi N}\,T_A(\vec{b})\,/\,A\right]^A}
{\sigma^{eff}_{\Psi N}\,A}\,,
\label{glaubpa}
\ee
being $T_A(\vec{b})$ the nuclear thickness function. 

\begin{figure}[t]
\centerline{\psfig{figure=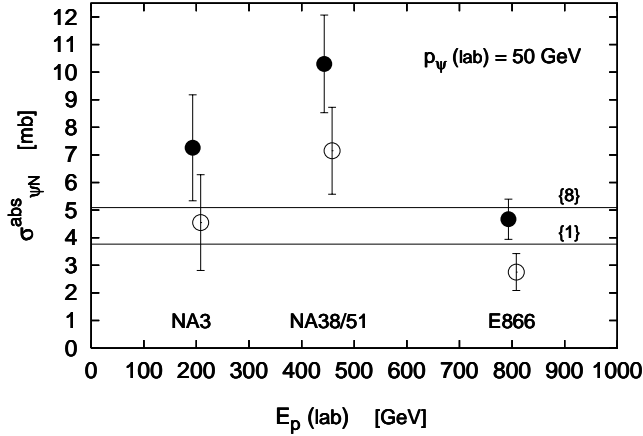,width=8.5cm}} 
\protect\caption{Results of the fits to pA data for the various experiments 
considered. Open dots refer to the Glauber model fit while full dots account 
for gluon anti-shadowing 
corrections. The horizontal lines indicate the theoretical expectations for the
singlet and octet absorption mechanisms, respectively.}
\label{protnucfit}
\end{figure}

For $A = 208$ and $p_\Psi = $ 50 GeV, using a realistic profile for Pb 
\cite{NUCDATA} and forcing the two expressions in 
eqs.$\,$(\ref{glaufirst}) and (\ref{glaubpa}) to be equal, we obtain 
$\sigma^{eff}_{\Psi N} = 3.8$ mb. This number has to be interpreted as the value 
of the effective absorption cross section of $\Psi$ on a color singlet nucleon,
including formation time effects. To obtain the corresponding value for the case
when $\Psi$ scatters on a color octet nucleon, according to the result of the 
previous section we have to enhance the singlet result by 35$\%$, obtaining 5.1 mb.
The values of 
$\sigma^{eff}_{\Psi N} = 3.8$ mb and 5.1 mb are indicated as full lines in
FIG.~\ref{protnucfit}. Because of a possible retardation of the
color mechanism, we expect the experimental values to lie between these two lines.

We have then re-analyzed the existing data taken by 3 fixed target 
experiments: NA3 at CERN \cite{NA3}, which measured
the ratio of $\Psi$ production cross sections between Pt and H$_2$ for 200 GeV
projectile protons; NA38/51 \cite{NA38/51} at CERN, which measured the cross
section for different targets (H$_2$, D$_2$, C, Al, Cu, W) struck by 450 GeV
protons; E866 at FERMILAB \cite{E866}, which measured the ratio of cross sections 
between W and Be for 800 GeV protons. In order remove the effects on absorption
due to different $\Psi$ momenta, we impose the same kinematic conditions, 
considering 
$\Psi$'s of 50 GeV momentum in the laboratory frame, corresponding to $x_F$ = 0.15,
0.01, $-$0.04 respectively, for the three experiments. We then assume that the 
absorption mechanism is parameterized by the effective cross section 
$\sigma^{eff}_{\Psi N}$, appearing in the Glauber model expression given in
eq.$\,$(\ref{glaubpa}). The ratio of cross sections is defined as
\be
R_{A/B} = \sigma_\Psi^{pA}\,\mbox{\large $/$}\,\sigma_\Psi^{pB}
\,.
\ee
Therefore, while the $pA$ cross section depends on the two parameters 
$\sigma_\Psi^{NN}$ and $\sigma^{eff}_{\Psi N}$, the ratio depends only on the 
latter. We made three separate fits using the MINUIT package from CERNLIB, including
quoted statistical and systematic errors for the various cross sections and ratios.
We extract the effective absorption cross section by minimizing $\chi^2$.
The result of these fits is presented in the second column of TABLE~\ref{pafit}. 
\begin{table}[b]
\caption{Results of the fits to pA data for the various experiments considered.
See text for details.}
\begin{center}
\begin{tabular}{|c||c|c|} \hline
Experiments & \multicolumn{2}{c|}{$\sigma^{abs}_{\psi N}$ [mb]} \\\cline{2-3}
$p_\psi$ = 50 GeV & Glauber fit & Gluon corr.   \\ \hline\hline
    NA3         &$\ $ 4.5 $\pm$ 1.7 $\ $&$\ $ 7.3 $\pm$ 1.9 $\ $ \\ \hline
    NA38/51     &$\ $ 7.1 $\pm$ 1.6 $\ $&$\ $ 10 $\pm$ 1.8 $\ $ \\ \hline
    E866        &$\ $ 2.7 $\pm$ 0.7 $\ $&$\ $ 4.7 $\pm$ 0.7 $\ $ \\ \hline\hline
    Averages    &$\ $ 3.5 $\pm$ 0.6 $\ $&$\ $ 5.7 $\pm$ 0.6 $\ $ \\ \hline\hline
 Theory & \multicolumn{2}{c|}{3.8 \{{\bf 1}\} $\ \ \ \ $ 5.1 \{{\bf 8}\}} \\ \hline
\end{tabular}
\end{center}
\label{pafit}
\end{table}
The average is obtained, using the weights $w_i = 1/\delta \sigma_i^2$,
as ${\overline \sigma} = \sum_i w_i\,\sigma_i / \sum_i w_i$, while the error is 
$\delta \sigma = (\sum_i w_i)^{-1/2}$. The values extracted from experiment are
represented as open points in FIG.~\ref{protnucfit}. The mean value of $3.5 \pm
0.6$ mb seems to favor the singlet mechanism, for which a cross section of 
3.8 mb is theoretically expected. 

On the other hand, the assumption that $\Psi$ is produced on a bound nucleon with
the same cross section as on a free one, used in
eq.$\,$(\ref{glaubpa}), can be an oversimplification of the physics under study.
We know that in a nuclear environment parton distributions are substantially 
modified \cite{GP96,FS99,EKS99}. In particular, the gluon distribution in nuclei
exhibits anti-shadowing effects at $x_{Bj} \simeq 0.1$, {\it i.e.} the gluon 
distribution $g_A(x_2,Q^2)$ in a nucleus is larger than the one in a nucleon 
$g(x_2,Q^2)$. We expect that charmonium production on a nucleus is enhanced by a 
factor $R^g_A(x_2,Q^2) = g_A(x_2,Q^2)/g(x_2,Q^2)$. We estimate these effects
by rescaling the $\Psi$ cross section with the A-dependent correction 
$R_A^g = 0.032 \log(A) + 1.006$, extracted from \cite{EKS99}. For a heavy nucleus 
this 
amounts to an enhancement of the production cross section by a factor of nearly 1.2,
therefore requiring a larger $\sigma^{eff}_{\Psi N}$. We fit the data
once more, obtaining the results listed in the third column of TABLE~\ref{pafit}.
One sees that the new average value of $5.7 \pm 0.6$ mb now seems to favor the
octet mechanism, for which we estimated a value of 5.1 mb. 
The results of the fits, together with the theoretical expectations, are all plotted
in FIG.~\ref{protnucfit}, where the full points are the result of accounting for
anti-shadowing. One observes that the situation is, in all, quite uncertain. 

The above analysis is based on the assumption that there is no dependence of nuclear
effects on the proton energy, is the charmonium energy is the same in all
experiments. This is justified in many phenomena, e.g. the coherence time effect
\cite{HKN96}. However, there might still be some dependence on the proton energy.
According to \cite{KN84}, energy loss of the projectile gluon in the nuclear medium
substantially reduces the production rate of charmonia at large $x_F$ and is still
rather important at small $x_F$. In order to compensate for energy
loss, the gluon distribution must be evaluated at a higher value of $x_1$, therefore
suppressing $\Psi$ production. This will in turn reduce the final state absorption on
nucleons in a fit to the data. To account for this effect, already observed in 
Drell-Yan data \cite{J00}, further study, which
goes beyond the scope of this paper, is required.

To summarize, the data seem to suggest that $\Psi$ is absorbed by colored nucleons
in pA collisions. The picture is though not yet conclusive due to the large 
experimental errors. Nevertheless, we follow our theoretical prediction and examine
the potential role of the color mechanism in the more complex case of $AB$
collisions.

\section{Color dynamics in nucleus-nucleus collisions}

Preparing the ground for the forthcoming discussion, it is useful to recall the 
conventional expression within the Glauber model for the $\Psi$ production cross
section at impact 
parameter $\vec{b}$. Generalizing eq.$\,$(\ref{glaufirst}), to the case of $AB$
collisions, one simply obtains
\bea
\frac{d^2\sigma_\Psi^{A B}}{d^2\vec{b}}(\vec{b}) & = & 
\!\!\int \!\!d^2\vec{b}_A\,dz_A\ d^2\vec{b}_B\,dz_B 
\ \delta^2(\vec{b}_A +\vec{b}_B - \vec{b}) \ \sigma_\Psi^{NN} \nonumber \\
& \times &
\rho_A(z_A,\vec{b}_A) \ \left[1 - \sigma^{abs}_{\Psi N}\,T_A(z_A,\vec{b}_A)/A
\right]^{A-1} \nonumber \\ 
& \times &
\rho_B(z_B,\vec{b}_B) \ \left[1 - \sigma^{abs}_{\Psi N}\,T_B(z_B,\vec{b}_B)/B
\right]^{B-1} \nonumber \\ 
& \times &
S_{F.S.I.}(z_A,z_B,\vec{b}_A,\vec{b}_B)\,,
\label{glauber}
\eea
which describes the collision of two rows at impact parameters $\vec{b}_A$ and
$\vec{b}_B$, overlapping due to the $\delta$-function. 
$\rho_A$ and $\rho_B$ are the nuclear densities while
$$
T_A(z_A,\vec{b}_A) = \int_{- \infty}^{z_A}\!\!\! dz'_A\, \rho_A(z'_A,\vec{b}_A) 
\ \ \ \mbox{and}
$$
$$
T_B(z_B,\vec{b}_B) = \int_{z_B}^{+ \infty}\!\!\!\! dz'_B\, \rho_B(z'_B,\vec{b}_B)
$$
are the thickness functions. The last factor 
$S_{F.S.I.}(z_A,z_B,\vec{b}_A,\vec{b}_B)$ represent the probability of survival 
due to any additional ``anomalous'' suppression factor which goes beyond
conventional nuclear effects,
that is final state interactions with the produced matter (QGP, hadron gas, etc.).
Here we are not primarily interested in these contributions, therefore  
set $S_{F.S.I.} = 1$. On the other hand, we focus on the reinterpretation of the
terms describing nuclear effects discussing, in the next two subsections,
color exchange in the context of multiple scattering, then applying
the obtained results to the calculation of $\Psi$ production in $AB$ collisions.

\subsection{Multiple color exchange in a nucleus}

\begin{figure}[t]
\centerline{\psfig{figure=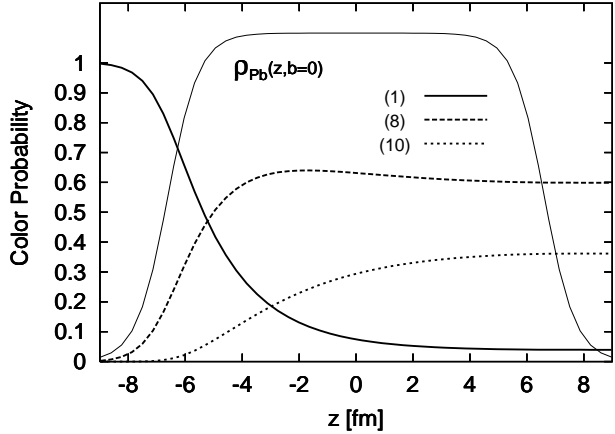,width=8.5cm}} 
\protect\caption{Evolution of the color probabilities $P_1$, $P_8$, 
$P_{10}$, along $z$ and with $\vec{b} = 0$. Also shown is the profile 
$\rho_{\mbox \small {\rm Pb}}(z,\vec{b}=0)$ of a Pb nucleus.}
\label{EVOLVE}
\end{figure}

Having completed the computation of the $\Psi$ cross sections with colored nucleons,
we need to look at how these color objects develop in a collision between nuclei.
Consider a nucleon of one of the nuclei colliding with a row of nucleons of the
other nucleus. It undergoes multiple scattering, with repeated color exchanges.
Keeping the projectile quark coordinates $\vec{y}_{1,2,3}$ frozen, according to the 
high-energy approximation, it is possible to calculate the probability
${\cal P}_n(\{\vec{y}_j\},z,\vec{b})$ for a nucleon in color state 
$n = \,${\bf 1},{\bf 8},{\bf 10} after having traveled through the row up to $z$ 
at impact parameter $\vec{b}$. This can be done in a very elegant way, by solving an
evolution equation for the color density matrix of the 3-quark system. The
problem has been considered some time ago with the general aim of studying the color
exchange interaction in the nuclear environment. A detailed calculation \cite{KL83} 
shows that
\be
{\cal P}_n = \frac{1}{27}\,\times
\left\{
\begin{tabular}{cc} 
$\ 1 + 20\ {\cal F} +\ 2\ {\cal G}$    
& $\ \ \ \ \{\mbox{\bf 1}\} $    \\ 
$16 - 40\  {\cal F} +\ 8\ {\cal G}$ 
& $\ \ \ \ \{\mbox{\bf 8}\} $    \\ 
$10 + 20\  {\cal F} - 10\ {\cal G}$    
& $\ \ \ \ \{\mbox{\bf 10}\} $    \\ 
\end{tabular}
\right.\,,
\label{probbor}
\ee
where
\be
{\cal F}(\{\vec{y}_j\},z,\vec{b}) = 
\exp\mbox{\Large $[$}-\mbox{\large $\frac{9}{16}$}\,\sum_{i<j} \sigma(y_{ij})
\ T(z,\vec{b})\,\mbox{\Large $]$}
\label{fcoeff}
\ee
and
\be
{\cal G}(\{\vec{y}_j\},z,\vec{b}) = \sum_{i<j}
\ \exp\mbox{\Large $[$}-\mbox{\large $\frac{9}{8}$}\, \sigma(y_{ij})
\ T(z,\vec{b})\,\mbox{\Large $]$}\,.
\label{gcoeff}
\ee
The coefficients ${\cal F}$ and ${\cal G}$ are expressed in terms of the standard
dipole cross section $\sigma(y_{ij})$, appearing in eq.$\,$(\ref{deltadip}), which 
one can construct by pairing two of the three quarks
in a nucleon. Their arguments are the differences of quark coordinates 
$\vec{y}_{ij} = \vec{y}_{j} - \vec{y}_{j}$. The thickness function $T(z,\vec{b})$ 
can be taken as that of nucleus $A$ or $B$, depending on which nucleus one is 
considering. We can average the probabilities with spatial nucleon wave functions as
\be
P_n(z,\vec{b}) = \mbox{{\large $\langle$} $\!\!N\!\!$ {\large $|$}}
\ {\cal P}_n(\{\vec{y}_j\},z,\vec{b})\ 
\mbox{{\large $|$} $\!\!N\!\!$ {\large $\rangle$}}\,,
\label{probavg}
\ee
therefore integrating over internal quark coordinates. Due to the
structure of eq.$\,$(\ref{probbor}), the averaging procedure expressed by 
eq.$\,$(\ref{probavg}) is transferred to an average of ${\cal F}$ and ${\cal G}$.
In analogy to the definition of $P_n(z,\vec{b})$, we will define also the 
functions $F(z,\vec{b})$ and $G(z,\vec{b})$. This average can in principle be
exactly evaluated, although in practice one needs to know the correct form
of the dipole cross section and of the spatial distribution of the three quarks
in a nucleon. We therfore choose to average the exponents appearing in
eqs.$\,$(\ref{fcoeff}) and (\ref{gcoeff}), instead of the full exponentials, {\it
i.e.} evaluating $\exp(-\langle N|X|N \rangle)$ instead of $\langle N| \exp(- X) |N
\rangle$ for a given $X$. This corresponds to neglecting inelastic shadowing 
corrections which, according to \cite{G69,M75}, are small. 
Due to rotational symmetry one has $\langle\,\sigma(y_{ij})\,\rangle = 
2/3\,\sigma^{in}_{NN}$ for all $i,j$. We took the value $\sigma^{in}_{NN} = 30$ mb.
In this way one obtains
\be
P_n(z,\vec{b}) \!=\! \frac{1}{27} \times\!
\left\{
\begin{tabular}{cc} 
$\ 1 + 20\ F(z,\vec{b}) +\ 6\ G(z,\vec{b})$    &  $\ \{\mbox{\bf 1}\} $   \\ 
$ 16 - 40\ F(z,\vec{b}) + 24\ G(z,\vec{b})$    &  $\ \{\mbox{\bf 8}\} $   \\ 
$ 10 + 20\ F(z,\vec{b}) - 30\ G(z,\vec{b})$    &  $\ \{\mbox{\bf 10}\} $    \\ 
\end{tabular},\right.
\label{probalb}
\ee
where
\be
\!\!\!F(z,\vec{b}) = \exp\left[- \mbox{\large$\frac{9}{8}$}\ \sigma^{in}_{NN}
\ T(z,\vec{b})\right]
\label{fcoeffalb}
\ee
and
\be
G(z,\vec{b}) = \exp\left[- \mbox{\large$\frac{3}{4}$}\ \sigma^{in}_{NN}
\ T(z,\vec{b})\right]\,.
\label{gcoeffalb}
\ee
One notices several properties of the probabilities. First of all, the
limit $T \rightarrow 0$ implies $F,G \rightarrow 1$. This means that 
$P_1 \rightarrow 1$ and $P_8,P_{10} \rightarrow 0$. In other words the nucleon,
before entering the nucleus, is in singlet state as it should be. Moreover,
$P_1 + P_8 + P_{10} = 1$ for any $T$, therefore probability is always conserved.
Finally, if the nucleus is large enough
one has $T \gg 1$ and $F,G \ll 1$. This implies that if sufficient scatterings 
have taken place, the color probabilities reach the statistical limit, given by the 
first coefficients of eqs.$\,$(\ref{probalb}). In other words, the 3-quark system
becomes unpolarized in color after many collisions. Out of 27 states, one is the
singlet, 16 are in the two octets and 10 are in the decuplet.
The $z$-dependence of the color probabilities is illustrated in FIG.~\ref{EVOLVE},
together with the longitudinal profile $\rho_{\mbox \small {\rm Pb}}(z,\vec{b})$ of
a Pb nucleus. Soon after the nucleon has penetrated the nuclear 
profile, a process that involves $\sim 4$ fm corresponding to 2 nucleon mean free
paths, the statistical limit is essentially reached in which about $16/27 \simeq 2/3$
of the nucleons are in octet state while $10/27 \simeq 1/3$
are in decuplet. The amount of singlet is negligible.

With the calculated probabilities $P_n$ and the previously evaluated color cross
sections $\sigma^{abs}_{\Psi N_n}$, we can now construct the cross section
for $\Psi$ when it scatters with a nucleon, taking into account all its color 
states. We obtain 
\be 
\sigma^{abs}_{\Psi N}(z,\vec{b}) = 
\sum_n \ \sigma^{abs}_{\Psi N_n}\ P_n(z,\vec{b})\,.
\label{weightedcs}
\ee
As obvious from FIG.~\ref{EVOLVE}, the projectile nucleon which collides with
several target nucleons quickly reaches a ``color equilibrium''. Using the values
for $\delta_n$ given in TABLE~\ref{psinuc}, that is 1.35 for the octet cross section
and 1.7 for the decuplet one, the effective absorption cross section of $\Psi$ on
a nucleon, within a nucleus-nucleus collision, becomes
\bea
\sigma^{abs}_{\psi N}(z,\vec{b}) & &
\ \mathop{\mathop{\longrightarrow}_{z \rightarrow \infty}}_{\vec{b} = 0} 
\ \left[\,\frac{1}{27} + \frac{16}{27} \times 1.35 + \frac{10}{27} 
\times 1.7\,\right] \times \sigma^{abs}_{\psi N} \nonumber \\
& & \ \ \ \simeq\ \ \ 1.47 \times \sigma^{abs}_{\psi N}
\ = \ 5.5 \ \mbox{mb}\,,
\label{asympt}
\eea
where the value $\sigma^{abs}_{\psi N} = 3.8$ mb has been used, according to our
calculation in the previous section.
We can therefore summarize saying that, although the cross section for $\Psi$ with
a singlet nucleon is 3.8 mb, including formation time effects, its effective value
in $AB$ collisions with large nuclei is 5.5 mb. 

As a final remark, notice that the calculation leading to eqs.$\,$(\ref{probalb}),
(\ref{fcoeffalb}), (\ref{gcoeffalb}) and (\ref{weightedcs}) is performed assuming
that each projectile nucleon undergoes collisions with singlet nucleons in the 
target. This neglects the color excitation of the target. A complete treatment
of the problem is very involved and is not expected to change the picture drawn. In
fact, as seen in FIG.~\ref{EVOLVE}, color excitation is a rather fast process and 
accounting for the neglected features would make it even faster. Moreover, as
discussed in the introduction, gluon radiation, here neglected, would speed
up color excitation even more.

\subsection{Charmonium Suppression in AB collisions}

\begin{figure}[t]
\centerline{\psfig{figure=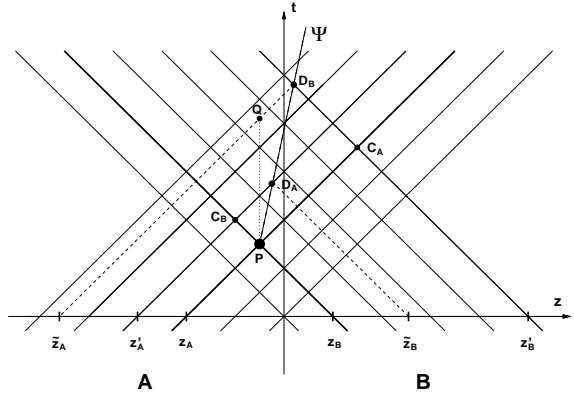,width=7.5cm,angle=-90}} 
\protect\caption{Longitudinal space-time representation of a row on row collision
taking place at impact parameters $(\vec{b}_A,\vec{b}_B)$ within a $AB$ collision.
The $\Psi$ production point is indicated with P, while the absorption points by
D$_{\mbox \small {\rm A}}$ and D$_{\mbox \small {\rm B}}$, do to nucleons of
$A$ and $B$ respectively.}
\label{nucnuc}
\end{figure}

Having discussed in detail colored cross section and multiple color exchange,
we obtained, with eq.$\,$(\ref{weightedcs}), the important ingredient needed
to calculate in detail the $\Psi$ production cross section in $AB$ collisions. 
It is first of all necessary to discuss some kinematical features concerning
$\Psi$ absorption and follow the longitudinal space-time 
diagram of FIG.~\ref{nucnuc}, where the kinematics is illustrated. Considering
the absorption caused by 
one of the nucleons of nucleus $B$, with coordinate ${z'}_{\!B}$, one notices that,
before interacting with the meson at point D$_{\mbox \small {\rm B}}$, it has 
traveled through nucleus
$A$ from $+ \infty$ down to ${\tilde z}_A$. The position ${\tilde z}_A$ depends on
the $\Psi$ velocity in the center of mass frame. This, in turn, depends on
$x_F = 0.15$, from which one obtains that
$v_\Psi = \sqrt{x_F^2\,s / (4m_\Psi^2 + x_F^2\,s)}$, where $s$ is the center of 
mass energy of a $NN$ collisions. With an elementary geometrical reasoning, making
use of the triangle $PQD_B$ for nucleus $B$, and $D_AC_BP$ for nucleus $A$, one 
arrives at the relations
\be
\tilde{z}_A = z_A - (z'_B - z_B)\,R_\Psi
\,, \ \ \tilde{z}_B = z_B - (z'_A - z_A)\,R_\Psi^{-1}\,,
\ee
where $R_\Psi = (1 - v_\Psi)/(1 + v_\Psi)$.
These must be used as arguments for the effective cross section in
eq.$\,$(\ref{weightedcs}). Therefore one has two cross sections, one for 
absorption by nucleus $A$, taking into account the multiple scattering of its 
nucleons with nucleus $B$, and one for nucleus $B$, depending on $A$. 
We can now modify eq.$\,$(\ref{glauber}) by making the replacements
\bea
\sigma^{abs}_{\Psi N}\ T_A(z_A,\vec{b}_A) & \rightarrow & \!\int_{- \infty}^{z_A} 
\!\!\!dz'_A\ \Sigma_B(\tilde{z}_B,\vec{b}_B)\ \rho_A(z'_A,\vec{b}_A)\,, \ \ \ \ \ \,\\
\sigma^{abs}_{\Psi N}\ T_B(z_B,\vec{b}_B) & \rightarrow & \!\int_{z_B}^{+ \infty} 
\!\!\!\!dz'_B\ \Sigma_A(\tilde{z}_A,\vec{b}_A)\ \rho_B(z'_B,\vec{b}_B)\,.\ \ \ \ \ \,
\eea
The position-dependent effective cross sections
\bea
\Sigma_A(\tilde{z}_A,\vec{b}_A) & = & \!\! \sum_{n} 
\ \sigma^{abs}_{\Psi\,N_n}\ P^A_{n}(\tilde{z}_A,\vec{b}_A)\,, \label{effA}\\
\Sigma_B(\tilde{z}_B,\vec{b}_B) & = & \!\! \sum_{n} 
\ \sigma^{abs}_{\Psi\,N_n}\ P^B_{n}(\tilde{z}_B,\vec{b}_B)\,, \label{effB}
\eea
are obtained from eq.$\,$(\ref{weightedcs}).

\begin{figure}[t]
\centerline{\psfig{figure=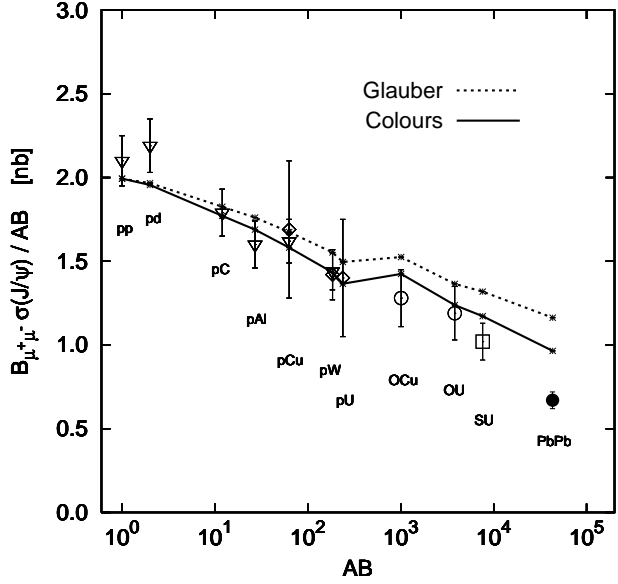,width=8.5cm}} 
\protect\caption{Calculations of total J/$\psi$ production cross sections in
$AB$ collisions and comparison with the NA38/50/51 data. The curves, drawn to guide
the eye, correspond to Glauber absorption (dashed) and to absorption by colored
nucleons (full).}
\label{total}
\end{figure}

Now eq.$\,$(\ref{glauber}) can be used to evaluate total cross sections, with
a further integration in impact parameter. We compute values to compare with all 
available data measured by the NA38/50/51 collaboration at the SPS \cite{NA5097}.
We use realistic
nuclear profiles \cite{NUCDATA} and scale the center of mass energy according to
the parameterization $\sigma_\Psi^{NN} \sim (1 - m_{J/\psi}/\sqrt{s})^{12}$.
The results are compared with the data in FIG.~\ref{total}.
The dashed curve corresponds to a standard Glauber model calculation with the
singlet absorption cross section $\sigma^{abs}_{\Psi N_1} = 3.8$ mb, as
evaluated in Sect.~3, while the thick curve refers to the improved calculation 
which takes into account color excitations, studied in Sect.~4. In order to 
emphasize the difference between the two absorption mechanisms, we neglect 
here anti-shadowing effects.
For the $pA$ calculation we increased the value of $\sigma^{abs}_{\Psi N_1}$ by
35$\,\%$, therefore used $\sigma^{abs}_{\Psi N_8} = 5.1$ mb. This is the extreme
situation, when $\Psi$ always encounters color octet nucleons. For the $AB$ case
we used eqs.$\,$(\ref{effA}) and (\ref{effB}). Since our main intent is to compare the
the new absorption mechanism to the conventional Glauber model one, we do not 
present the calculation of the $E_\perp$-dependent cross section for Pb+Pb 
collisions, which we cannot explain.

Different from the Glauber model result, our improved one provides a better agreement 
with the data, except for the Pb+Pb point. In other words, the proposed absorption 
mechanism due to color dynamics is found to be an important contribution
to nuclear effects in $\Psi$ production. It must be accounted for, together with
other mechanisms of suppression \cite{HK98,HuHK00,HKP} which we here put aside.
In the search for new physics one must in fact rely on a solid baseline, employing 
the best of our knowledge of the dynamics of $\Psi$ absorption at the early
stage of a nuclear collision.

\section{Conclusions}

In this work we studied the dynamics underlying $\Psi$ suppression
due to nuclear effects, employing the idea that color exchange processes in multiple
$NN$ collisions may play a significant role.

First of all, we calculated inelastic cross sections for $\Psi$ scattering on
colored nucleons and found an increase of about $35\,\%$ when the nucleon is in a color
octet state as compared with the singlet case, while the enhancement factor for
the color decuplet is about $70\,\%$.

Making use of available calculations of $\Psi N$ cross sections, we
included formation time effects and obtained the effective value for  $\Psi$ 
absorption on singlet nucleons $\sigma^{abs}_{\Psi N_1} = 3.8$ mb. Scaling this
value with a $35\,\%$ increase we obtained $\sigma^{abs}_{\Psi N_8} = 5.1$ mb
and with a $70\,\%$ increase $\sigma^{abs}_{\Psi N_{10}} = 6.4$ mb.

We then performed a fit to available data on $\Psi$ production cross sections
in $pA$ collisions. The result of the fit, including gluon anti-shadowing 
corrections, is that the effective absorption cross section for $\Psi$ is
$\sigma^{eff}_{\Psi N} = 5.7 \pm 0.6$ mb, therefore suggesting that in $pA$ 
collisions, the produced charmonium interacts with color octet nucleons. The
conclusion is, though, rather weak because the experimental points are quite
scattered.

Total production cross sections for $\Psi$ were then calculated and compared in
FIG.~\ref{total} with available data for $AB$ collisions. Compared to the Glauber
model result, the agreement is improved and we conclude that effects due 
to color exchange processes are
substantial and provide the value $\sigma^{abs}_{\Psi N} = 5.5$ mb for the
effective absorption cross section on nucleons in $AB$ collisions (See 
eqs.$\,$(\ref{weightedcs}) and (\ref{asympt})). 

In conclusion, we have completed a study of nuclear effects in charmonium production
in $AB$ collisions, providing a better baseline for future calculations. These amount
to an improved treatment of gluon bremsstrahlung, including non-linearities arising
because of gluon cascades, to the calculation of effects caused by gluon energy loss
and to the evaluation of $E_\perp$-dependent cross sections. This will hopefully 
provide the starting point from which to address more precisely the nature of 
anomalous suppression in Pb+Pb collisions.

\section*{Acknowledgments}

This work has been supported in part by the B.M.B.F. under contract number 06 HD 642.


\end{document}